\documentclass[%
 aip,
 amsmath,amssymb,
 nofootinbib,
 reprint,%
]{revtex4-1}

\usepackage{graphicx} 
\usepackage{natbib} 
\usepackage[usenames,dvipsnames]{color} 
\usepackage{amsmath} 
\usepackage[urlcolor=blue, hyperindex, colorlinks, bookmarks=true,linkcolor=black,citecolor=black]{hyperref} 
\usepackage{dcolumn}
\usepackage{amssymb} 
\usepackage{soul} 
\usepackage{ifthen} 
\usepackage{bbm}
\usepackage{comment}
\usepackage[ampersand]{easylist}
\usepackage{todonotes}
\usepackage{upgreek}
\usepackage{todonotes}

\usepackage{lipsum}

\def\be#1\ee{\begin{equation}#1\end{equation}}
\def\ba#1\ea{\begin{align}#1\end{align}}
\def\bg#1\eg{\begin{gather}#1\end{gather}}
\def\t{\text}

\def\shownote{1} 
\newcommand{\note}[1]{\ifthenelse{\shownote=1}{\textcolor{red}{[[#1]]}}{}}

\def\shownoteal{1} 
\newcommand{\nal}[1]{\ifthenelse{\shownoteal=1}{\textcolor{blue}{[[#1]]}}{}}
\def\shownotenj{1} 
\newcommand{\nnj}[1]{\ifthenelse{\shownotenj=1}{\textcolor{brown}{[[#1]]}}{}}

\begin{document}

\title{Aluminum air bridges for superconducting quantum devices realized using a single step electron-beam lithography process}

\author{N. Janzen}
\altaffiliation{Authors to whom correspondence should be addressed:\\N. Janzen, ncjanzen@uwaterloo.ca; \\A. Lupascu, alupascu@uwaterloo.ca}
\affiliation{Institute for Quantum Computing, Department of Physics
and Astronomy, and Waterloo Institute for Nanotechnology, University
of Waterloo, Waterloo, ON, Canada N2L 3G1}

\author{M. Kononenko}
\affiliation{Institute for Quantum Computing, Department of Physics
and Astronomy, and Waterloo Institute for Nanotechnology, University
of Waterloo, Waterloo, ON, Canada N2L 3G1}

\author{S. Ren}
\affiliation{Institute for Quantum Computing, Department of Physics
and Astronomy, and Waterloo Institute for Nanotechnology, University
of Waterloo, Waterloo, ON, Canada N2L 3G1}

\author{A. Lupascu}
\altaffiliation{Authors to whom correspondence should be addressed:\\N. Janzen, ncjanzen@uwaterloo.ca; \\A. Lupascu, alupascu@uwaterloo.ca}
\affiliation{Institute for Quantum Computing, Department of Physics
and Astronomy, and Waterloo Institute for Nanotechnology, University
of Waterloo, Waterloo, ON, Canada N2L 3G1}

\date{\today}

\begin{abstract}
In superconducting quantum devices, air bridges enable increased circuit complexity and density as well as mitigate the risk of microwave loss arising from mode mixing. We implement aluminum air bridges using a simple process based on single-step electron-beam gradient exposure. The resulting bridges have sizes ranging from 20~$\upmu$m to 100~$\upmu$m, with a yield exceeding 99~\% for lengths up to 36~$\upmu$m. When used to connect ground planes in coplanar waveguide resonators, the induced loss contributed to the system is negligible, corresponding to a loss per bridge less than $1.0\times10^{-8}$. The bridge process is compatible with Josephson junctions and allows for the simultaneous creation of low loss bandages between superconducting layers.   

\end{abstract}

\maketitle

Superconducting circuits have emerged as one of the primary candidates for the implementation of quantum computing~\cite{kjaergaardSuperconductingQubitsCurrent2020, krantzQuantumEngineerGuide2019}. One of the primary advantages of superconducting circuits is the flexibility in design and scalability resulting from their realization as microfabricated circuits on a chip. As the complexity and scale of superconducting circuits increases, microfabrication methods that enable increased density and quantum coherence are essential. Relevant approaches for scalability include multi-chip fabrication~\cite{rosenberg3DIntegratedSuperconducting2017, kosenBuildingBlocksFlipchip2022, foxenQubitCompatibleSuperconducting2017, goldEntanglementSeparateSilicon2021} and three dimensional wiring integration~\cite{bejaninThreeDimensionalWiringExtensible2016, rahamimDoublesidedCoaxialCircuit2017}.

Air bridges are one of the most relevant components of high-complexity superconducting circuits. They are used as crossovers to enhance functionality by enabling increased circuit density and improved coherence in both distributed resonators and qubits through the prevention of mode mixing~\cite{ponchak_excitation_2005}. Wirebonds, which are an alternative crossover to air bridges, are ineffective in prevention of mode mixing~\cite{chen_fabrication_2014}. Several methods have been developed to date to implement air bridge crossovers for superconducting circuits. These processes are carried out using techniques including using a photo resist scaffold followed by etching of the excess aluminum to separate the bridge~\cite{lankwarden_development_2012, chen_fabrication_2014}, employing resist stacks that use different resists to define both a scaffold as well as a resist window above the scaffold with an undercut for liftoff~\cite{abuwasib_fabrication_2013,jin_microscopic_2021}, and forming a dielectric scaffold that is etched after Al deposition~\cite{dunsworth_low_2018}. In this Letter, we present a method to fabricate aluminum air bridges that uses a simple single-step fabrication process that relies on gradient exposure electron-beam lithography (EBL). This process produces bridges with lengths ranging from 20~$\upmu$m to 100~$\upmu$m, with a yield exceeding 99~\% for bridges of lengths up to 36~$\upmu$m. We characterize coplanar waveguide resonators with integrated air bridges and find that negligible microwave loss is introduced when these bridges are used to connect ground planes. The process is compatible with fabrication of Josephson junctions for superconducting  qubits and it has the additional desirable feature that it allows for fabrication of bandage-style contacts between different superconducting layers~\cite{dunsworth_characterization_2017}.
\begin{figure*}[!]
\includegraphics[width=6.0in]{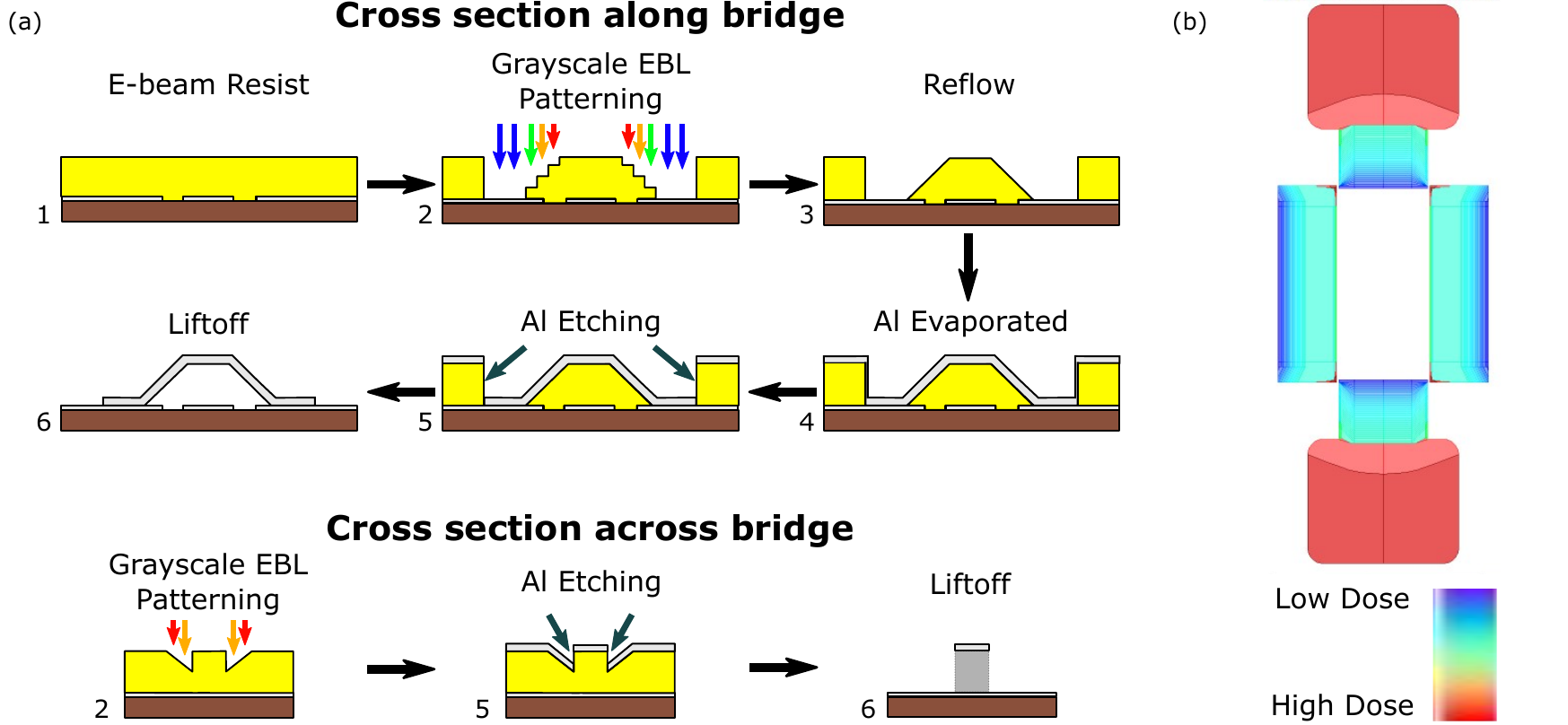}
\caption{\label{fig:Process} a)The process key steps, showing cross sectional views along the bridge (top) and across the bridge (bottom). The step numbers in the two views correspond, with only a subset of the steps for the view across the bridge. The substrate, resist, and aluminum layers are shown in brown, yellow, and gray respectively. b)Dose pattern used to define gradient exposure for a suspended bridge.}
\end{figure*}

We start by first presenting the key elements of the process, shown schematically in Fig.~\ref{fig:Process}(a). The starting configuration is a chip that contains the relevant metallic layers, suitably patterned, that require connection through air bridges. Next, a layer of electron-beam resist is spun, with a thickness corresponding to the target elevation of air bridges. A gradient exposure pattern using EBL, followed by development and reflow, is used to create a smooth resist profile. This forms the support of the bridge, with a sharp separation from the top surface of the unexposed resist. After evaporation of Al and liftoff of the resist, air bridges are formed, which connect relevant points of the metallic base circuit. We note that grayscale based exposure for formation of metallic air bridges were explored in previous work~\cite{girgis_fabrication_2006,papageorgiou_one-step_2013}. Here we develop a single step air bridge process that is integrated into existing superconducting circuit processes. 

\begin{figure*}[!]
\includegraphics[width=6.8in]{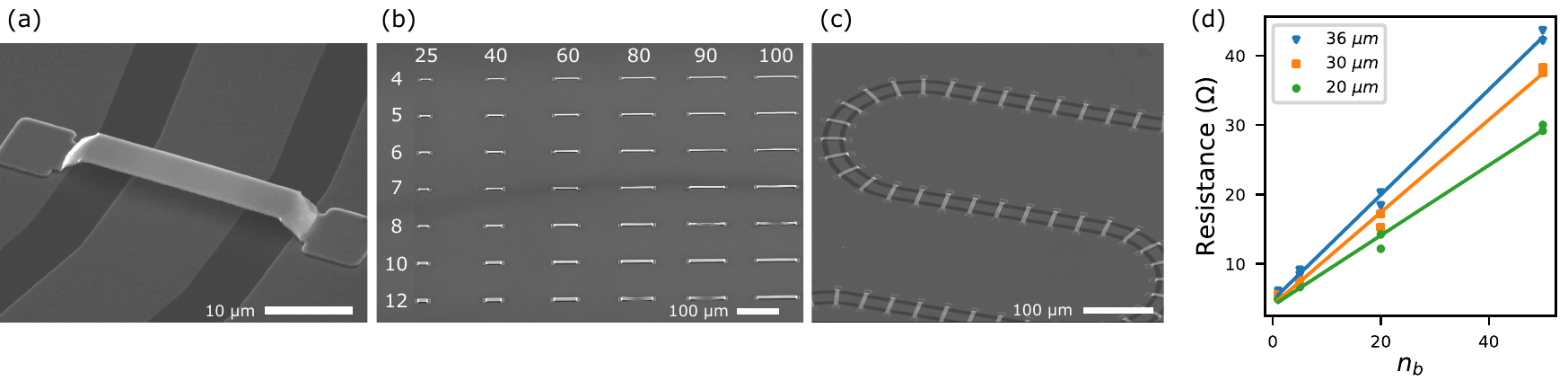}
\caption{\label{fig:BridgesAndResistance} a)SEM picture with a tilted angle of a bridge with a length of 36~$\upmu$m and a width of 6~$\upmu$m. b)SEM pictures of bridges of varying sizes, ranging in length from 25~$\upmu$m to 100~$\upmu$m and width from 4~$\upmu$m to 12~$\upmu$m. c) SEM picture showing an array of bridges of length 36~$\upmu$m separated by 40~$\upmu$m along a coplanar waveguide.  d) Corrected resistance for bridge arrays of versus the number of bridges $n_b$. The bridges have a width of 6~$\upmu$m, a thickness of 300~nm, and lengths of 20~$\upmu$m, 30~$\upmu$m, and 36~$\upmu$m.}
\end{figure*}

Next, we discuss our complete fabrication process, which is formed of four main layers. The substrate is p-doped high-resistivity silicon (resistivity larger than 10~{k}$\Omega\cdot$cm). \emph{Layer 1 - alignment markers.} Alignment markers for subsequent lithography steps are defined using dry etching through an S1811 resist mask, patterned with optical lithography, which is stripped after etching with sonication in Remover PG at $80^\circ$C. The substrate is further cleaned through sonication in acetone followed by IPA. \emph{Layer 2 - base circuit layer.} The base circuit layer is used to define circuit components including ground planes, waveguides, and capacitor pads. It is made with negative lithography using a layer of ma-N 1410 resist with a thickness of 1275~nm, followed by evaporation of 100~nm of aluminum and liftoff in Remover PG. \emph{Layer 3 - Josephson junctions.}  Josephson junctions are generated by a standard double-angle evaporation process using a resist bi-layer consisting of 110~nm of PMMA 950K A3 on 410~nm of PMGI, followed by evaporating 40~nm and 70~nm of Al at $\pm20^\circ$ respectively. \emph{Layer 4 - air bridges.}  Three layers of PMMA 950K A6 are spun at 1000 RPM onto the sample to create a 3~$\upmu$m resist stack. Hard baking at $180^\circ$C is performed for 1/1/3~min for each layer respectively. Grayscale EBL is performed using the dose map shown in Fig. \ref{fig:Process}(b) with a development in MIBK:IPA 3:1 solution for 180~s followed by selective reflow on a hotplate at $110^\circ$C for 5~min. Next, argon milling, performed at an angle normal to the surface with voltage and current settings 400~V and 30~mA respectively, is used to etch the circuit aluminum layer to enable low-resistance contacts. A 450~nm thick layer of Al is evaporated onto the scaffold. The sample is etched for 80~s in Transene Al etchant Type A to improve bridge separation before lifting off the excess metal in Remover PG at $80^\circ$C for 45~min. The sample is directly transferred to another bath of Remover PG for 15~min, followed by acetone and IPA cleanings before lightly blow drying with nitrogen. In Fig.~\ref{fig:BridgesAndResistance}(a) a scanning electron microscope picture (SEM) of a typical bridge is shown. The bridges were also examined at a large angle to confirm the complete removal of the PMMA scaffold under the bridge.

In the course of process development, the optimization of some key parameters is essential for achieving a high process yield. Optimization of the bridge dose map, including consideration of proximity exposure effects, is critical to ensure flat bridge surfaces and create the necessary separation for liftoff. We use selective reflow~\cite{schleunitz_selective_2011} to smooth the incline of the bridge to increase its strength. Selective reflow relies on the fact that partially exposed PMMA has a lower glass transition temperature than unexposed resist. Baking the sample at $110^\circ$C for 5~min allows the partially exposed PMMA at the bridge incline to reflow, while the unexposed sidewalls of the bridge pads remain vertical. Further, we include an Al etching step to remove any residual Al that builds up on the vertical sidewalls of the bridge profile during evaporation. While these build-ups are very thin and did not impact yield, a brief Al etching step removes these without any etching the previous metal layers.

The process can be used to produce bridges with lengths ranging from 20~$\upmu$m to 100~$\upmu$m. This flexibility is demonstrated in Fig.~\ref{fig:BridgesAndResistance}(b) which shows an SEM images of bridges over a wide range of dimensions. We note that bridges with a length of 36~$\upmu$m are sufficiently long to connect ground planes across coplanar waveguides with a signal line width of 16~$\upmu$m and a gap width of 8~$\upmu$m (as shown in Fig.~\ref{fig:BridgesAndResistance} (c)), which are typical in the fabrication of high quality factor resonators (Q~$>10^5$). Thus, air bridges with length up to 36~$\upmu$m are characterized to ensure they function as effective crossovers. To measure electrical resistance, a test is performed where arrays of $n_b$ bridges of a size suitable for coplanar waveguides (length 20-36~$\upmu$m) are fabricated to measure the yield and the resistance at room temperature across the entire array (see Fig.~\ref{fig:BridgesAndResistance}(d)) The bridges are connected in series via Al wires deposited prior to the bridges which are subtracted from the resistance based on a measurement of the sheet resistance of this wire. The resistance corresponding to bridges shows a linear scaling with the number of bridges, with a best fit that is within 1\% of the value calculated from the sheet resistance. There is a 4.3~$\Omega$ offset in the resistance, which is due to the measurement setup, including connecting wires and contact pads. The measurements (estimates based on sheet resistance) are 0.506 (0.503)~$\Omega$, 0.667 (0.665)~$\Omega$, and 0.758 (0.760)~$\Omega$ for bridges of length 20~$\upmu$m, 24~$\upmu$m, 36~$\upmu$m respectively. Four-probe probe measurements show a negligible contact resistance between the bridge pads and the base aluminum layer of 7~m$\Omega$, suggesting an Ohmic contact. This result confirms uniformity and is an indicator that there are no open contacts. These arrays are also an indicator of high yield for bridges with length up to 36~$\upmu$m. Based on optical inspection and the resistance measurements, all 456 bridges of this devices were successful providing a yield with a lower bound of 99.2~\% within a 95~\% confidence interval. Bridges with larger dimensions have a lower yield. Fig.~\ref{fig:BridgesAndResistance}(b) shows examples of longer bridges that have collapsed. While the yield for longer bridges was not quantified in detail, it is clearly lower than for the bridges up to 36 um length. We expect that the yield of longer bridges can be improved in future work using a critical point dry step to reduce unwanted stresses on the bridge from surface tension and N2 pressure endemic to blow-drying that may cause bridge collapses.

\begin{figure}
\includegraphics[width=3.4 in]{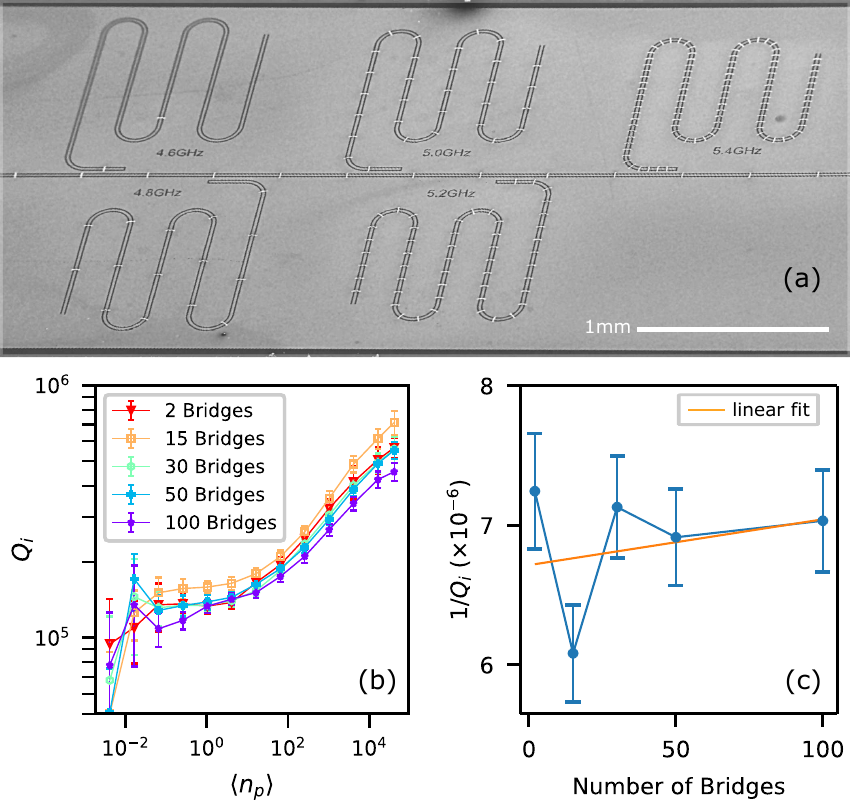}
\caption{\label{fig:Resonators} a)SEM picture with a tilted angle of a device formed of five resonators coupled to a common feedline. b)Quality factor versus average photon population $\langle{}n_p\rangle$ for each resonator (dots, with lines as guide to the eye). The error bars correspond to the uncertainty in the parameter extracted from the fit of $S_{21}$ around each resonant frequency. The number of bridges fabricated on the resonators range from 2 to 100. c) Loss (inverse quality factor) versus number of bridges in the single photon regime ($\langle{}n_p\rangle\sim 1$) where the system enters the desaturated limit (dots) and a linear fit (continuous line).}
\end{figure}

Next, we discuss experiments in which the quality factor of coplanar waveguide resonators, fabricated with air bridges, is characterized at low temperatures. The device, shown in Fig.~\ref{fig:Resonators}(a), consists of five quarter-wavelength coplanar waveguide resonators coupled to a common feedline with the number of bridges on the resonators ranging from 2-100. The resonators are designed to have resonance frequencies distributed uniformly between 4.6~GHz and 5.4~GHz. The large separation between the resonance frequencies allows for the measurement of transmission through each resonator to be done in a muliplexed way. Experiments are performed in a dilution refrigerator, at a temperature of 30~mK. The transmission through each resonator, after subtraction of the background transmission, is given by~\cite{gao_physics_2008,deng_analysis_2013}
\begin{align} \label{eq:S21}
   S_{21}(\omega)=\frac{1+2iQ_i\frac{\omega-\omega_0}{\omega_0}}{1+\frac{Q_i}{Q_e}+i\frac{Q_i}{Q_\alpha}+2iQ_i\frac{\omega-\omega_0}{\omega_0}}, 
\end{align}
where $\omega$ is the probe frequency, $\omega_0$ is the resonant frequency, $Q_i$ is the internal quality factor, $Q_c$ is the external quality factor resulting from capacitive coupling to the feedline, and $Q_\alpha$ is a parameter that characterizes the asymmetry of $S_{21}$ that arises due to transmission line non-idealities~\cite{deng_analysis_2013}. Figure~\ref{fig:Resonators}(b) shows the internal quality factor versus average photon population, $\langle{}n_p\rangle$, after fitting the model in Eq. \ref{eq:S21}. For each resonator, the internal quality factor has a minimum at low powers corresponding to the loss due to two-level-system defects in the desaturated limit~\cite{mullerUnderstandingTwolevelsystemsAmorphous2019}, which occurs when the photon population is at single photon levels ($\langle{}n_p\rangle\sim 1$). The single photon regime quality factors, ranging between 130k and 160k, are relatively modest compared to optimized high quality factor aluminum resonators~\cite{megrantPlanarSuperconductingResonators2012, earnestSubstrateSurfaceEngineering2018a}. While we minimize possible contamination and ensure the exposure to air is only a few hours during packaging, we emphasize that our process does not incorporate important elements such as vigorous substrate cleaning and optimized etching based processes for the base aluminum layer~\cite{earnestSubstrateSurfaceEngineering2018a} required for fabrication of high-quality factor aluminum resonators, given that the focus in this work is the development and optimization of the bridge process. However, the dependence of the intrinsic quality factor on number of bridges, shown in Fig.~\ref{fig:Resonators}(c), gives a quantitative indicator of the microwave loss associated with each bridge. We note that the bridges are uniformly distributed along the resonator, and therefore both electric field (voltage) and magnetic field (current) related loss mechanisms are captured. In addition, we expect that mode conversion loss is effectively mitigated for all the resonators~\cite{ponchak_excitation_2005,chen_fabrication_2014}. With an assumed linear dependence of loss on the number of bridges, our experiments allow us to estimate a loss per bridge with an upper bound of $1.0\times10^{-8}$. This is an excellent figure of merit, given that it is expected that several bridges at the scale of a resonator are sufficient to mitigate slotline modes \cite{ponchak_excitation_2005} and improve ground plane uniformity.

\begin{figure}[!]
\includegraphics[width=3.4 in]{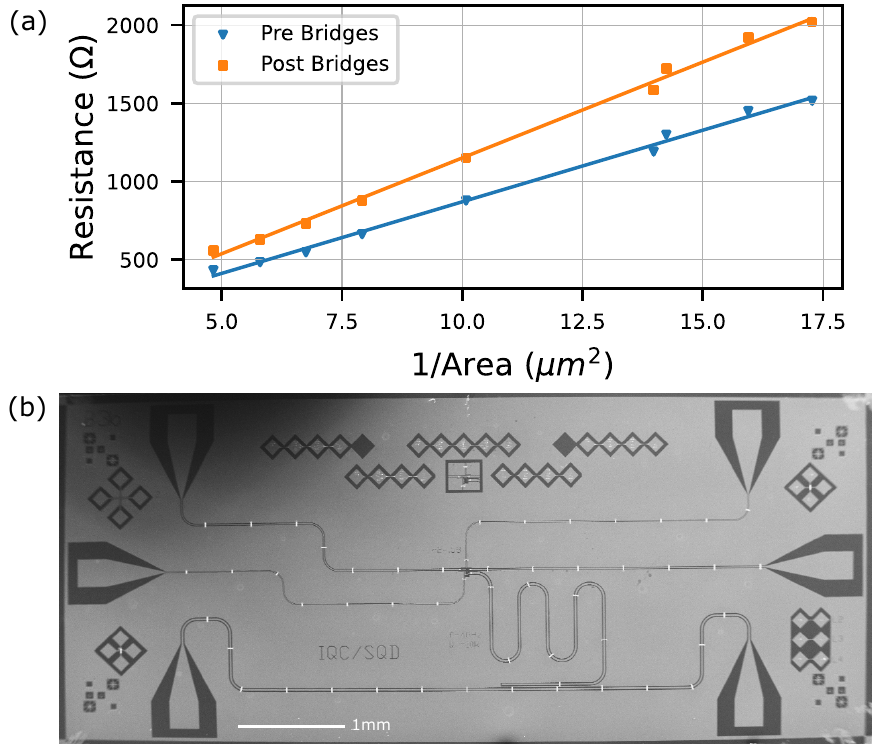}
\caption{\label{fig:JosephsonJunctions} a)Resistance of individual Josephson junctions plotted as a function of the inverse overlap area, measured before and after inclusion of the bridge layer. b)SEM picture of a device for light-matter interaction experiments.}
\end{figure}

The integration of air bridges with high-quality Josephson junctions in the same process is essential for fabrication of superconducting circuits including qubits. We chose a process implementation where the Josephson junctions are fabricated in a layer prior to the air bridges. The processing of the chip required to make the bridges is found to have an effect on the resistance of Josephson junctions. In Fig.~\ref{fig:JosephsonJunctions}(a) we plot the resistance of a set of junctions versus inverse area, for areas ranging from 0.06~$\upmu\t{m}^2$ to 0.21~$\upmu\t{m}^2$ before and after bridge processing. The dependence is linear and it shows a systematic and reproducible increase in resistance of approximately $30\%$ that arises from the 5~min of total bake time at $180^\circ$C, consistent with previous studies~\cite{migacz_thermal_2003}. We note that the estimated critical current density after the bridge process is 2.6~$\upmu \t{A}/\upmu\t{m}^2$, compatible with flux qubits~\cite{yurtalan_characterization_2021} and larger than typical critical current densities used in transmons. The junctions are unaffected by other steps of this process, including the argon milling step and etching steps, as they are protected by the 3~$\upmu$m coat of PMMA. A device for controlled light-matter interactions experiments implemented with a flux qubit coupled to a transmission line has been recently realized using this process (see Fig.~\ref{fig:JosephsonJunctions}(b)) \cite{Janzen_Coupler}. This device contains six Josephson junctions and its operation confirmed expected target critical current values for these junctions. These results are promising in view of integration of the bridge process with coherent devices. One interesting feature of our process is that the bridge layer can also be used to create bandage connections between Josephson junctions and other circuit layers~\cite{dunsworth_characterization_2017}, eliminating the need of an additional processing step.

In conclusion, we demonstrated an effective EBL air bridge process for superconducting quantum devices. In contrast to previous work, our process has a simple implementation based on a single-layer EBL gradient exposure. The process has a high yield and has the capability to fabricate bridges with sizes from 20~$\upmu$m to 100~$\upmu$m, allowing for a wide range of applications. We found that the induced microwave loss is negligible and the process has a reproducible effect on the Josephson junction resistance. Further work will focus on the investigation of the properties of bridges in the presence of currents and on optimizing the contacts for high critical currents, as well as realization of high coherence devices.

\begin{acknowledgments}
 We are grateful to Ali Yurtalan for advice during the early stages of the project and to Xi Dai for commenting on the manuscript. We acknowledge support from NSERC through Discovery and RTI grants, Canada Foundation for Innovation, Ontario Ministry of Research and Innovation, and Industry Canada. We would like to acknowledge support from CMC Microsystems and Canada’s National Design Network (CNDN). We would like to thank the staff of the University of Waterloo's QNFCF for assistance during fabrication and in particular Greg Holloway who provided insight on many technical problems. The University of Waterloo's QNFCF facility was used for this work. This infrastructure would not be possible without the significant contributions of CFREF-TQT, CFI, ISED, the Ontario Ministry of Research \& Innovation and Mike \& Ophelia Lazaridis.
\end{acknowledgments}

\subsection*{Author contributions} \textbf{Noah Janzen:} Conceptualization (equal); formal analysis (lead); funding acquisition (supporting); investigation (lead); methodology (lead); project administration (supporting); software (equal); visualization (lead); writing – original draft (equal); writing – review \& editing (equal).
\textbf{Michal Kononenko:} Conceptualization (supporting); formal analysis (supporting); investigation (supporting); methodology (supporting); writing – review \& editing (supporting).
\textbf{Shaun Ren:} Formal analysis (equal); software (equal); visualization (supporting).
\textbf{Adrian Lupascu:} Conceptualization (equal); formal analysis (equal); funding acquisition (lead); investigation (supporting); methodology (supporting); project administration (lead); resources (lead); supervision (lead); visualization (supporting); writing – original draft (equal); writing – review \& editing (equal).


\section*{References}


%

\end{document}